# Ultrasmooth single crystal diamond surfaces resulting from implantation and lift-off processes

T. N. Tran Thi[*,a], B. Fernandez[a], D. Eon[a], E. Gheeraert[a], J. Härtwig[b], T. A. Lafford[b], A. Perrat-Mabilon[c], C. Peaucelle[c], P. Olivero[d] and E. Bustarret[a]

a. Institut Néel, CNRS and Université Joseph Fourier, Grenoble, France.

b. European Synchrotron Radiation Facility (ESRF), Grenoble, France.

c. Service Faisceaux d'Ions de l'Institut de Physique Nucléaire de Lyon, Université Claude Bernard Lyon 1, Lyon, France.

d. Faculty of Mathematics, Physics and Natural Sciences, University of Torino, Italy.

*A method for obtaining a smooth, single crystal diamond surface is presented, whereby a sacrificial defective layer is created by implantation and graphitized by annealing before being selectively etched. We have used $O^+$ at 240 keV, the main process variables being the ion fluence (ranging from $3\times10^{15}$ cm$^{-2}$ to $3\times10^{17}$ cm$^{-2}$) and the final etching process (wet etch, $H_2$ plasma and annealing in air). The substrates were characterized by atomic force microscopy, optical profilometry and white beam X-ray topography. The influence of the various process parameters on the resulting lift-off efficiency and final surface roughness is discussed. An $O^+$ fluence of $2\times10^{17}$ cm$^{-2}$ was found to result in sub-nanometre roughness over tens of μm$^2$.*

## 1. Introduction

The interest in boron-doped diamond for electronic devices is increasing because of its attractive electrical properties: wide band gap, high thermal conductivity, high mobility, and high breakdown electric field. Recently, nanostructuring diamond monocrystalline plates into 2D (delta-doped layers, membranes) or 1D objects (pillars, suspended cantilevers) is becoming increasingly relevant for a number of applications of this material (e.g. Field Effect Transistors, Sensors, Nano electro-mechanical system). In particular, the δ-doped diamond structure has been considered for some time because of its possible application as an electrical switch in Field Effect Transistors (FET), expected to commute high power at high frequency. A δ-doped diamond structure requires a very thin metallic boron-doped p$^+$ layer ($[B]_{p+} \geq 5\times10^{20}$ at.cm$^{-3}$), called the "δ-layer", intercalated between two low boron-doped layers (Non-intentionally Doped, NiD) ($[B]_{NiD} \leq 1\times10^{17}$ at.cm$^{-3}$).

In previous studies [1], it has been shown that the single crystal diamond substrates play a very important role for the quality of the epitaxial growth and that they also have a significant influence on the electrical properties of the devices which are subsequently fabricated. In the case of δ-layers, the roughness requirements on the individual surfaces and interfaces translate into specifications of the starting substrates that become quite stringent at the nanometric scale. These specifications led us in this study to develop a method of smoothing the single-crystal diamond surface down to the nanometric scale while maintaining a superficial structural quality compatible with defect-free epitaxial growth. In this work, we report the technique of surface smoothing using implantation and a subsequent lift-off process using either hot acid or hydrogen plasma, followed by annealing in air.

## 2. Single crystal diamond substrate

Nowadays, many types of single crystal diamond substrates are available on the market, such as Ib or IIa High Pressure and High Temperature HPHT or IIa, IIIa Chemical Vapour Deposition CVD, among others. For our study, we required a substrate with small misorientation, high crystalline quality (low defect density) and, in particular, the smoothest possible surface. However, most of the current diamond substrates available at a reasonable price (either CVD or HPHT-grown) have been mechanically polished, leaving a strongly anisotropic root mean square (rms) roughness. In some cases (either HPHT or CVD), deep holes are present, of a few 100 nm in diameter and depth. These are associated with polishing damage and emerging bundles of dislocations. Given this situation, we started our study with Ib HPHT (100) diamonds, 3×3×0.5mm$^3$, produced by Sumitomo Electric Co., Japan.

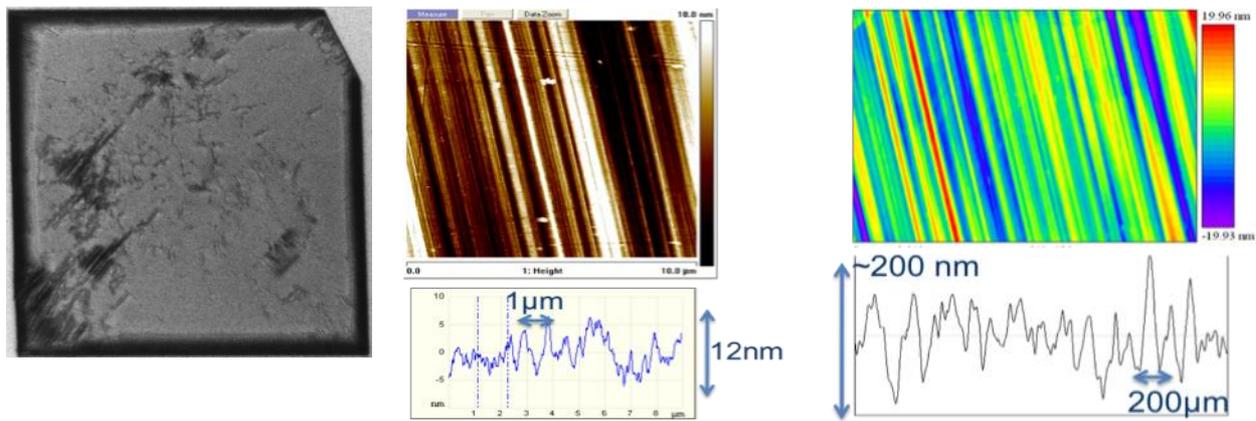

**(a)** Bulk crystal quality imaged by white beam X-ray topography. Entire sample imaged.

**(b)** AFM 10×10 µm² showing relatively low local roughness.

**(c)** Optical profile over 1.25×1.64 mm² showing greater roughness over longer length scales.

**Figure 1** Bulk and surface quality of a 100-oriented Ib HPHT Sumitomo 3×3×0.5mm³ substrate.

Such a standard Ib HPHT diamond normally has a misorientation of less than 1° and looks yellow to the eye due to nitrogen incorporation. The white beam X-ray topography image (Fig. 1a) taken at BM05, ESRF, clearly showed the defects such as dislocations in the bulk crystal. Properly oriented defects show up as dark or light contrast on the uniform grey background of more perfect regions. The dislocation density was found to be significant, up to $10^6$ cm$^{-2}$.

The as-delivered surface was studied using an Atomic Force Microscope (AFM) over a 10×10 µm² area (Fig. 1b) and an optical profiler for larger areas (Fig. 1c, 1.25×1.64 mm²). Both of these measurements showed a homogenous surface with parallel polishing lines of a few nm width and a spacing of period ~1 µm superimposed on parallel polishing "waves" (200 µm period). The local roughness along a small scan line was $r_{rms}$ ~ 4 nm while $R_{rms}$ ~ 40 nm on a mm length scale. The same result was obtained on both sides of the substrate. We needed therefore to develop a polishing method to remove the mechanical polishing-induced damaged top layer (as least 200 nm) and improve the roughness to less than 1 nm, in order to obtain an ultra-smooth surface on the diamond substrate. Although it may also lead to sub-nanometre roughness, we discarded mechanical polishing because in fragile crystals such as diamond it creates many subsurface extended defects.

Many non-mechanical surface smoothing processes have been proposed in the past for diamond, such as various kinds of plasma etching, ion beam milling or ion implantation and lift-off. We have tried several reactive ion etching techniques, such as Oxygen Electron Cyclotron Resonance etching, Oxygen Radio Frequency etching or Microwave etching in an $O_2$ + $SF_6$ + Ar gas mixture. However, under standard conditions, the roughness resulting from these plasma etching methods after removal of 200 nm of material were not significantly better than before. This led us to work on the latest method: ion implantation followed by a lift-off process.

### 3. Experimental results

Applying implantation to remove a layer of diamond is not a new idea. It was first introduced in 1992 [2] in order to remove a thin diamond film from a diamond substrate by implantation of $C^+$ or $O^+$ ions with energy 4 MeV to 5 MeV at fluences ranging from $6\times10^{16}$ cm$^{-2}$ to $1\times10^{18}$ cm$^{-2}$ with lift-off of the damaged zone by hot acid etching. The idea was then continued by [3-5] but in most cases used implantation and lift-off were meant to fabricate microstructure devices on diamond. The roughness of the apparent surface was generally not characterized, although recent results [6] reported a roughness value of 0.8 nm rms. Ion straggling has long been known to smooth out the surface roughness [2]. The whole process consists of four steps:

1) Ion implantation to create a buried damaged layer,

2) Annealing to transform the amorphous damaged layer into graphite,

3) Selective etching followed by lift-off to remove the graphite (and possibly the top surface layer),

4) Final annealing/etch to smooth the new surface and remove residual implantation damage.

First of all, the depth profiles for the ion range and damage range were calculated using the TRIM software [7]. The depth of the damaged layer, which is a buried region extending above the ion end-of-range depth (see Fig. 2a), is determined by the ion species and energy. In order to remove ~250 nm from the surface, we used $O^+$ implantation at an energy of 240 keV. The most important parameter determining the success of this step is the ion fluence. In this work, the implantations were performed with $O^+$ ions at 240 keV and ion fluences of $5\times10^{13}$, $3\times10^{14}$, $1\times10^{15}$, $5\times10^{15}$, $5\times10^{16}$, $1\times10^{17}$, $2\times10^{17}$, and $3\times10^{17}$ cm$^{-2}$. Before implantation, one corner of each sample was covered by an aluminum mask in order to distinguish the non-implanted regions from the implanted ones. After implantation, the diamond plates became dark (darker with increasing fluence). The samples were then annealed at 950°C for 2 hours in vacuum in order to transform the amorphous damaged layer into graphite. This step is considered to heavily damage the buried layer up to the end of the ion range [3]. The diamond plates remained dark after removal from the furnace, due to the damaged, implanted layer. Raman analysis was performed using a He-Ne laser (excitation wavelength: 632.8 nm). The measured Raman spectrum of a non-implanted area shows the optical phonon diamond peak at 1332 $cm^{-1}$. By contrast, the diamond peak is absent in the implanted area, replaced by large peaks at 1170 $cm^{-1}$ and 1250 $cm^{-1}$ which are associated with the disordered graphitized phase.

To remove the damaged layer (third step), we employed two methods: i) standard lift-off by hot acid $H_2SO_4+HNO_3+HCLO_4$ 1:1:1 at 150°C for more than 2 hours and ii) $H_2$ plasma at 950°C with -150 V applied bias voltage for 2 hours. When the implantation fluence was high enough, we successfully obtained a damaged layer which could be removed by both of these methods. Lift-off was achieved by implantation of oxygen ions at 240 keV with a fluence above $1\times10^{17} cm^{-2}$. First of all, we noted that the diamond plates returned from dark to bright yellow, which is the colour of the pristine diamond. Secondly, by optical profilometry, we observed clearly the step between the non-implanted and implanted zones. This step was 280 nm high - compare this with the peak in the simulated damage profile (at 230 nm) and ion range (at 250 nm), (see Fig. 2a).

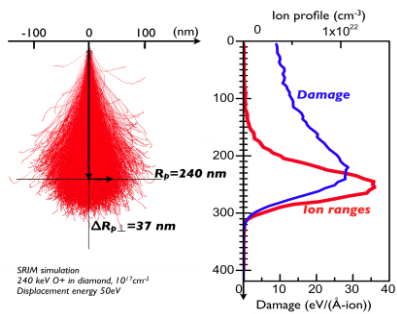

(a) TRIM simulations of the depth profile of the ion and damage ranges.

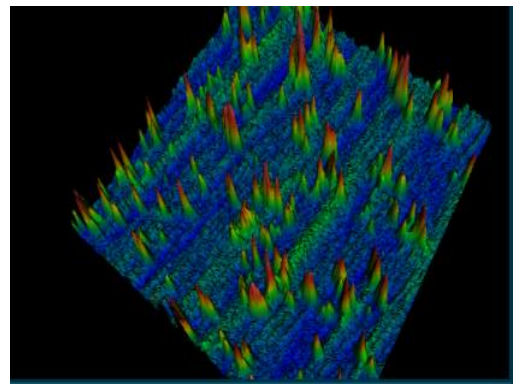

(b) Optical profile of a 50×50 μm$^2$ area on the diamond surface after the lift-off process.

**Figure 2** Oxygen implantation 240 keV: simulation and experiment

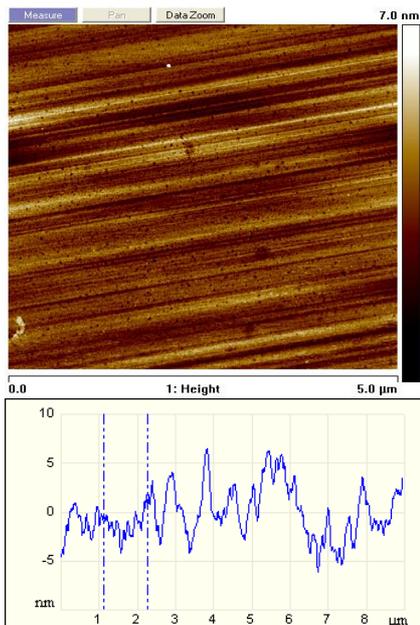

(a) Pristine diamond surface, rms roughness ~4nm.

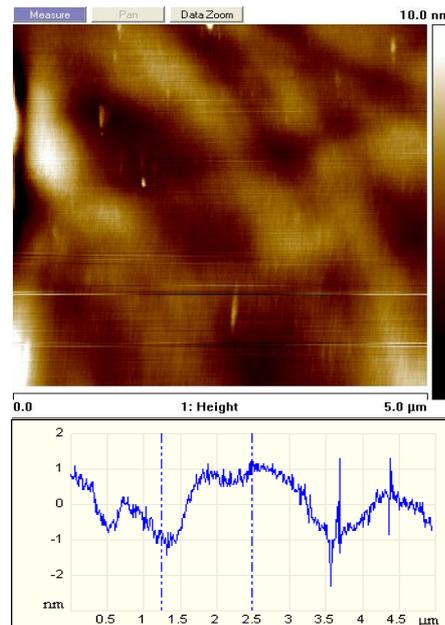

(b) Diamond surface after our smoothing process, rms roughness ~0.6nm.

**Figure 3** AFM images of a 5μm×5μm area of a Ib HPHT substrate before (a) and after (b) smoothing.

Following the lift-off process, optical profile images were taken in several areas of 50×50 μm$^2$ over the whole implanted surface (Fig.2b). The surface was smoother than before treatment, with an average roughness of 1 nm, compared with 4 nm previously measured. However, the surface was covered with 20 nm-high protuberances. The density of these spike-like defects was ~2×10$^4$ cm$^{-2}$. To remove them, we annealed the sample at 550°C in air for 2 hours. This provided a slow isotropic dry-etch at a rate of 100 nm/hr. Overall, ~500 nm were thus removed from the top surface of the sample. The AFM topography of this new surface is shown in Fig. 3. A first observation was that all the spikes defects were smoothed out. Moreover, the polishing lines disappeared and the diamond surface became very flat. The average rms roughness is 0.6 nm over 25 μm$^2$, in good agreement with recent results obtained with a similar process [6]. We have further checked that such air annealing processes did not improve the roughness of as-delivered (i.e non-implanted) substrates.

Fig. 4 shows the dependence of the surface rms roughness $r_{rms}$ on the ion fluence for the two lift-off processes described above. It is clear that the hydrogen plasma indeed improved the surface, but not sufficiently to reach sub-nm rms roughness. However, by using oxygen ion implantation at high fluence above 1×10$^{17}$ at.cm$^{-2}$ combined with either hydrogen plasma or hot-acid lift-off, we were able to remove a 500 nm thick top layer from the surface, without creating additional damage by mechanical polishing. We thus obtained a homogenous, flat surface without any polishing lines or morphological defects.

### 4. Conclusion

We have demonstrated that a diamond surface can be smoothed down to sub-nanometre rms roughness by lifting off a 500 nm top layer by a technique based on ion implantation. The present process starts with oxygen implantation, followed by annealing in vacuum at 950°C. A hydrogen plasma or an acid etch resulted in lift-off, and the final anneal in air at 500°C provided an additional soft isotropic etch. Lift-off only occurred for ion fluences above 1×10$^{17}$ cm$^{-2}$. The surface thus revealed was smooth at the sub-nanometre level over 5×5 μm$^2$ areas.

**Acknowledgement**

The authors would like to thank to J E.Butlet, J.Morse and J.Pernot for many valuable discussions, and acknowledge support from ANR project *"Deltadiam" – ANR – 08 – BLAN – 0195.*

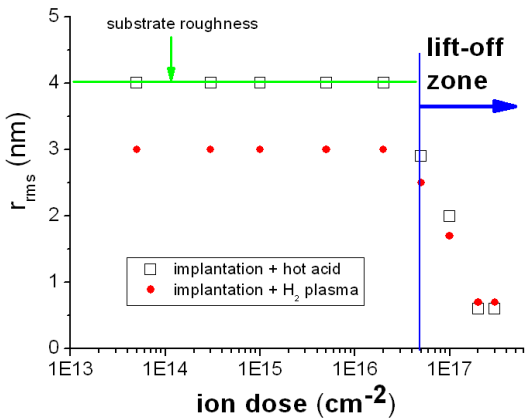

**Figure 4** Roughness of diamond face as a function of ion fluence for both lift-off process (hot acid and hydrogen plasma).

**References**

[1] D. Takeuchi *et al* Phys Rev B, Vol 63, 245328 (2011)

[2] Parikh N. R. *et al.* Appl. Phys. Lett.61 (26) (1992)

[3] Olivero P. *et al.* Diamond and Related Materials 15 1614 (2006)

[4] C. F. Wang *et al.* J. Vac. Sci. Technol. B 25 (3) (2007)

[5] Y. Mokuno *et al.* Diamond and Related Materials (2008)

[6] C. Mer-Calfati *et al.* Phys. Stat. Solidi (a) 206, 1955 (2009)

[7] J. F. Ziegler, Transport of Ions in Matter (TRIM), IBM Corp. software (1991)